\documentclass[journal]{IEEEtran}
\ifCLASSINFOpdf
\else
\fi
\usepackage{amsmath}
\usepackage{hyperref}
\usepackage{tikz}
\usepackage{pgfplots}
\usepackage{xcolor}
\usepackage{multirow}
\usepackage{subcaption}
\usepackage{url}
\usepackage{tikz}
\newcommand\copyrighttext{%
  \footnotesize This work has been submitted to the IEEE for possible publication. Copyright may be transferred without notice, after which this version may no longer be accessible}
\newcommand\copyrightnotice{%
\begin{tikzpicture}[remember picture,overlay]
\node[anchor=south,yshift=10pt] at (current page.south) {\fbox{\parbox{\dimexpr\textwidth-\fboxsep-\fboxrule\relax}{\copyrighttext}}};
\end{tikzpicture}%
}
\begin{document}
%
\title{UWB TDoA Error Correction using Transformers: Patching and Positional Encoding Strategies
}
%
%
%

\author{Dieter Coppens, Adnan Shahid, \IEEEmembership{Senior member , IEEE}, Eli De Poorter
\thanks{D. Coppens, A. Shahid and E. De Poorter are with IDLab, Department of Information  Technology, Ghent University—imec, 9052 Ghent,
Belgium (e-mail: dieter.coppens@ugent.be)}
}

%
%

\markboth{Journal of \LaTeX\ Class Files,~Vol.~14, No.~8, August~2015}%
{Shell \MakeLowercase{\textit{et al.}}: Bare Demo of IEEEtran.cls for IEEE Journals}
%



\maketitle
\copyrightnotice
\begin{abstract}
Despite their high accuracy, UWB-based localization systems suffer inaccuracies when deployed in industrial locations with many obstacles due to multipath effects and non-line-of-sight (NLOS) conditions. In such environments, current error mitigation approaches for time difference of arrival (TDoA) localization typically exclude NLOS links. However, this exclusion approach leads to geometric dilution of precision problems and this approach is infeasible when the majority of links are NLOS. To address these limitations, we propose a transformer-based TDoA position correction method that uses raw channel impulse responses (CIRs) from all available anchor nodes to compute position corrections. We introduce different CIR ordering, patching and positional encoding strategies for the transformer, and analyze each proposed technique's scalability and performance gains. Based on experiments on real-world UWB measurements, our approach can provide accuracies of up to 0.39 m in a complex environment consisting of (almost) only NLOS signals, which is an improvement of 73.6 \% compared to the TDoA baseline. 
\end{abstract}


%
\IEEEpeerreviewmaketitle

\section{Introduction}
\IEEEPARstart{R}{obust} and accurate indoor localization is increasingly important for several emerging applications such as healthcare, emergency services, smart logistics, sports tracking, and various location-based services \cite{surveyindoorlocapplications,smartlogistics}. Following this trend, Ultra-Wideband (UWB) technology has seen a surge in interest and is considered one of the more promising technologies for indoor positioning systems (IPS). UWB-based IPS can achieve centimeter-level positioning accuracy due to the use of large bandwidth and the accompanying short time duration of pulses. These signal characteristics enable higher temporal resolution and greater resilience against multipath effects \cite{coppens2022overview}, which is a significant advantage, when compared to traditional narrow-band radio techniques such as WiFi, 4G, 5G, BLE, etc.
\begin{figure}[h]
     \centering
     \begin{subfigure}[b]{0.4\linewidth}
         \centering
         \includegraphics[width=\linewidth]{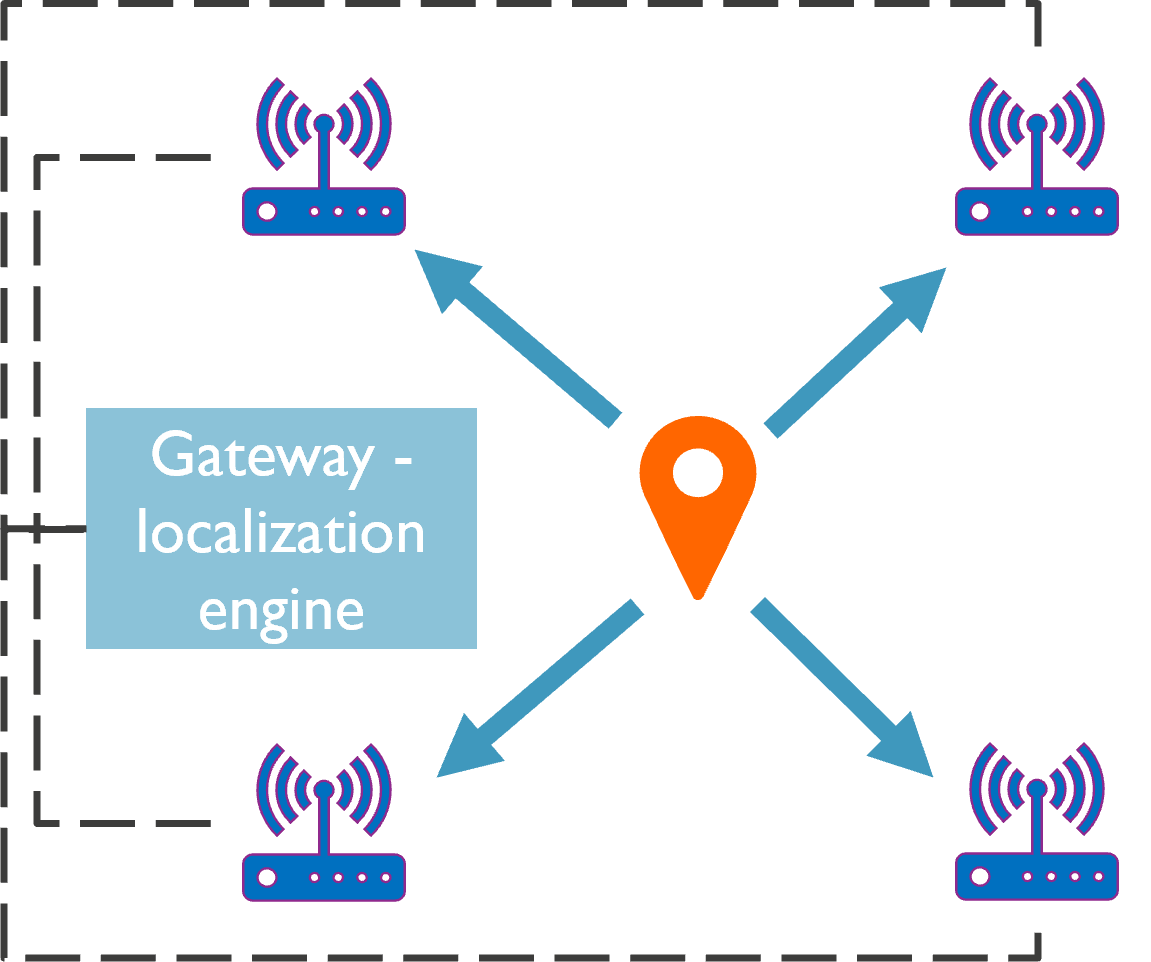}
         \caption{TDoA}
         \label{fig:tdoa1}
     \end{subfigure}
     \hfill
     \begin{subfigure}[b]{0.35\linewidth}
         \centering
         \includegraphics[width=\linewidth]{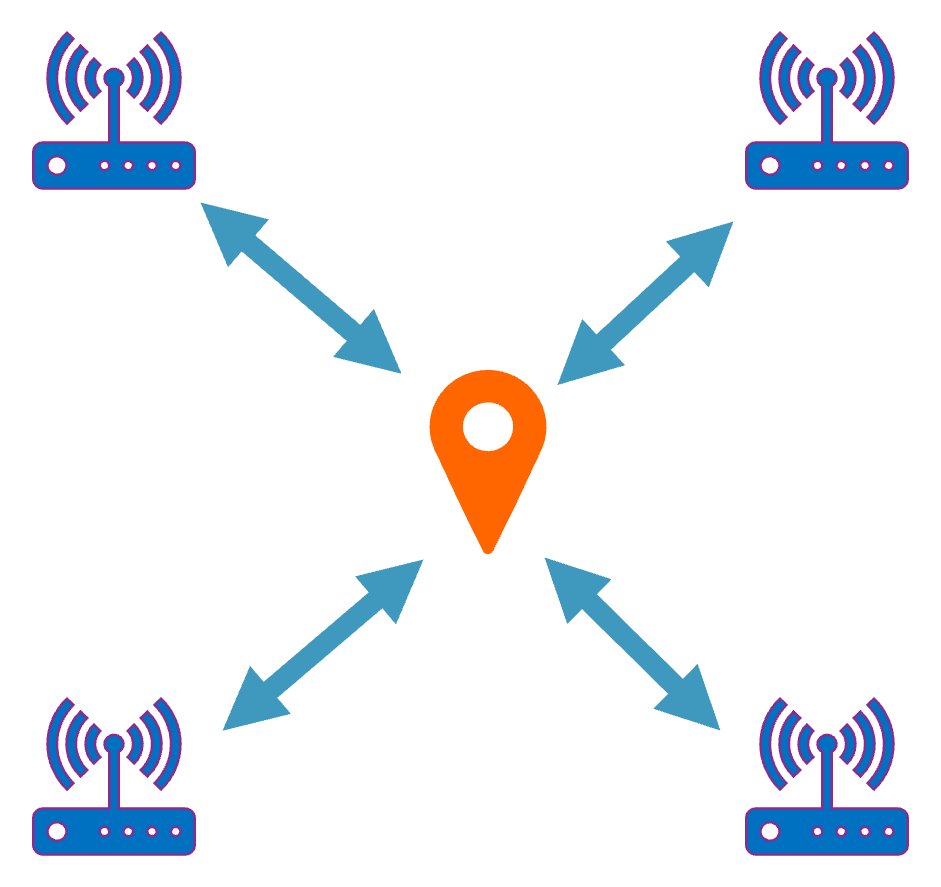}
         \caption{TWR}
         \label{fig:twr}
     \end{subfigure}
        \caption{Comparison of TDoA and TWR localization techniques}
        \label{fig:three graphs}
\end{figure}
Various localization techniques leverage the precise timing resolution of UWB to achieve accurate position calculations. Two-Way Ranging (TWR) determines distances by measuring the time of flight between the tag and the anchor, offering robustness against clock drift errors and removing the need for synchronization between the tag and the anchor. However, three packets are required to be sent between each tag-anchor pair. Time Difference of Arrival (TDoA) estimates positions by calculating the time difference in signal arrival across multiple anchor pairs leading to multiple differences in distance of arrival (DDoA). Each DDoA defines a hyperbolic curve representing possible tag positions. By using optimization techniques the most likely position can be found. TDoA requires only a single UWB transmission resulting in low power consumption for the tag but requiring clock synchronization between the anchors. In real-world scenarios, a major remaining challenge is the multipath behavior in non-line-of-sight (NLOS) conditions. UWB performance degrades in NLOS conditions due to blocked direct paths, introducing positive bias in range estimates. Indoor environments frequently experience NLOS scenarios from obstacles like walls, pedestrians, furniture and equipment \cite{20}. Moreover, multipath propagation can lead to stronger reflected signals than the attenuated direct path, further deteriorating positioning accuracy. Thus, mitigating NLOS and multipath effects is crucial for improving UWB IPS performance.

For TWR, there exist several works that provide error correction in NLOS conditions using machine learning. These methods are trained in a supervised way using datasets of UWB ranges and raw physical data like the Channel Impulse Response (CIR) \cite{mao2018probabilistic,jaron,li2023variational} and recently even self-supervised approaches have been proposed that do not require the collection of labeled datasets \cite{coppensselfsupervised}. 

In contrast, for TDoA, no error correction approaches have yet been proposed. Instead, these TDoA approaches fully remove non-reliable links by assessing the quality of the UWB communication channel using the CIR or other PHY features. However, recent work \cite{duong2024errormitigationtdoauwb} has shown that excluding larger numbers of UWB measurements can result in an error increase. This indicates that removing UWB measurements from TDoA should be done with caution: throwing away data, even from low-quality transmission, can worsen the performance. These techniques potentially discard valuable information, like multipath components, that could provide critical positioning insights. Hence, a method that can handle multipath signals without discarding potentially useful data is necessary. 

To overcome the limitations of existing approaches, we propose a novel TDoA position correction method based on a transformer encoder that leverages the CIRs from all anchor nodes to compute position corrections in a single pass directly.

The main contributions of this work are:
\begin{itemize}
    \item The first UWB TDoA error correction approach that corrects - rather than removes - TDoA signals based on raw CIR information 
    \item Design of a novel transformer encoder architecture capable of processing raw CIRs directly that is more scalable than traditional Convolutional Neural Network (CNN) approaches.
    \item Introduction of a novel encoding technique, encoding the 3D position of anchors, which is better suited for localization than traditional sequence positions as used for encoding in classical transformers.
    \item Analysis of the influence of multiple patching, positional encoding, and CIR ordering approaches on the error correction performance.
    \item Comparison of our approach with prior approaches in terms of both accuracy and complexity, showing our solution outperforms existing solutions.
\newline
\end{itemize}

The remainder of the paper is organized as follows. Section \ref{sec:relatedwork} discusses the related work for UWB error correction. In Section \ref{sec:problem} the UWB TDoA error correction challenge is mathematically modeled. Next, in Section \ref{sec:experiment} the environment in which the experiments are performed and the gathered datasets are described. Section \ref{sec:methodology} discusses the proposed methodology and Section \ref{sec:results} discusses the performance of the proposed architectures. Finally, the conclusion is given in Section \ref{sec:conclusion}.
\begin{figure*}[h!]
    \centering
    \includegraphics[width=0.7\linewidth]{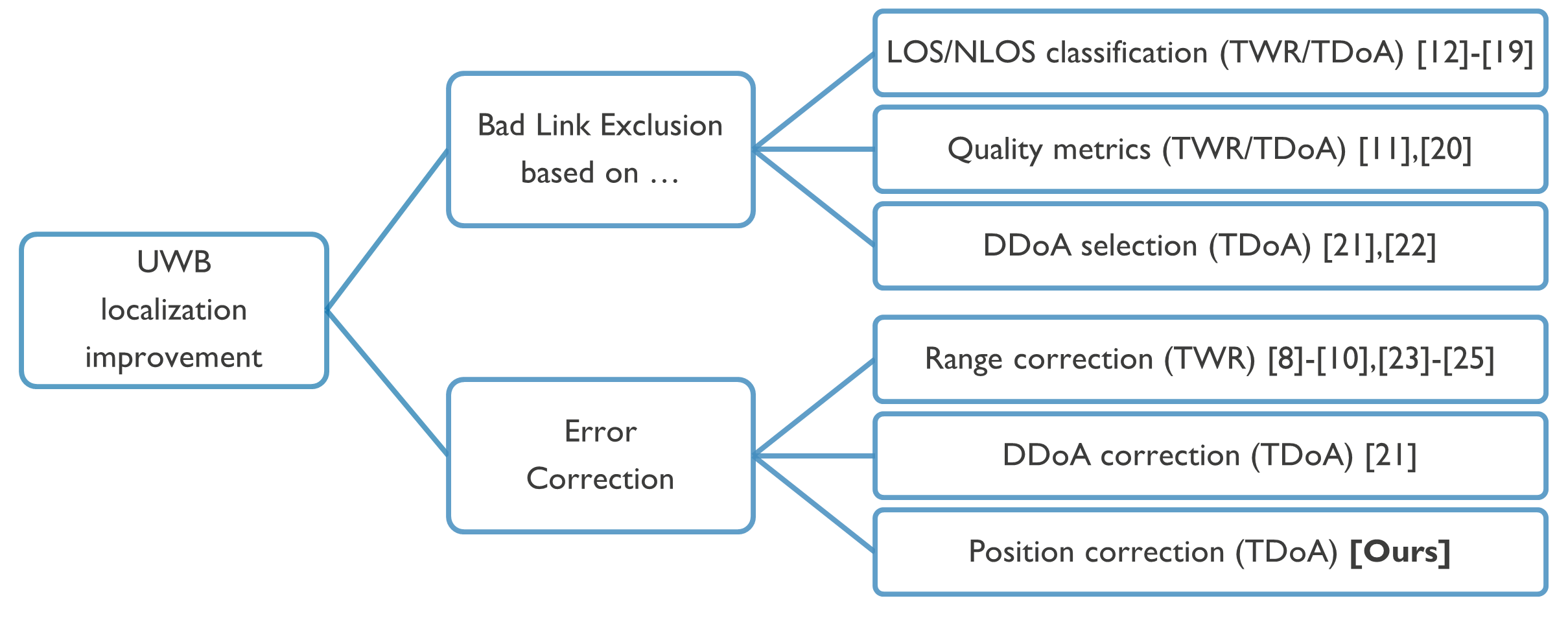}
    \caption{Overview of the different UWB localization improvement approaches, showing different techniques and how our work related to prior work.}
    \label{fig:relatedwork}
\end{figure*}

\section{Related Work}
\label{sec:relatedwork}
\begin{table*}[ht]
\centering
\caption{Comparison of our proposed approach with related work on TDoA improvement in LOS/NLOS environments. The table mentions the goal, the approach, the technology, the environment type, the number of anchors and the inputs that are used.}
\label{tab:relatedwork}
\begin{tabular}{c|ccccccc}
\textbf{Paper} & \textbf{Goal} & \textbf{Approach} &\textbf{Technology} &\textbf{Environment} & \begin{tabular}{@{}c@{}}\textbf{Uncorrected} \\ \textbf{TDoA MAE}\end{tabular} & \textbf{\#Anchors} & \textbf{Input} \\ \hline
\cite{zhao2019select} & DDoA correction & Non-convex optimization & / & Simulation & / & 21 & DDoA \\
\cite{ma2007nonline} & Bad link exclusion & Distribution functions & Cellular & Simulation & 70 m & 7 & \begin{tabular}{@{}c@{}}Hyperbola \\ intersections \end{tabular} \\
\cite{van2021anchor}& DDoA correction/selection & CNN & UWB & Experimental & 0.46 m & 8 & CIRs \\
\cite{duong2024errormitigationtdoauwb}& Bad link exclusion & Unsupervised clustering & UWB  & Experimental & 0.45 m  & 15 & CIRs \\
\textbf{Our work} &\textbf{Position correction} & \textbf{Transformer} & UWB & \textbf{Experimental}& \textbf{1.48} m& \textbf{15} & \textbf{CIRs} 
\end{tabular}
\end{table*}

This section provides an overview of the different methods proposed in the literature to improve UWB localization. We categorize the existing approaches into two primary groups: \textit{Bad Link Exclusion} and \textit{Error Correction}. Each category is further divided into specific techniques illustrated in Figure \ref{fig:relatedwork} and discussed in the subsections below.
\subsection{Bad Link Exclusion}
Bad link exclusion methods aim to identify and exclude unreliable links to improve the accuracy of UWB localization. These methods can be broadly categorized as follows:
\subsubsection{\textbf{LOS/NLOS Classification (TWR/TDoA)}}
The following papers try to classify each transmission as LOS or NLOS and based on this they exclude bad (NLOS) links. For TWR, removing NLOS links eliminates biased range measurements from the position calculation, allowing the remaining LOS links to provide accurate distances, thereby improving overall positioning accuracy. However, it reduces the geometric diversity and number of available measurements. For TDoA, excluding NLOS UWB measurements prevents corrupted time difference measurements from distorting the hyperbolic positioning solution, ensuring that only reliable LOS-based time differences contribute to the final position estimate. However, less receiver redundancy can lead to poor geometric dilution of precision. However, since each hyperbola is calculated based on timing information from two anchor nodes, removing a single anchor quickly reduces the number of available hyperbola. As such, in some areas of an environment, there might be insufficient LOS links to still calculate the position, resulting in the lack of any positioning estimate.
Early methods for NLOS identification relied on statistical models. For instance, \cite{che} used Generalized Gaussian Distribution (GGD) algorithms, while \cite{fan2019non} employed Gaussian mixture models (GMM). Later research explored classic machine learning classifiers, \cite{sangnlos} analyzed Support Vector Machines (SVM), Random Forests (RF), and Multilayer Perceptrons (MLP) to distinguish between LOS and NLOS conditions. More recent work has shifted towards deep learning, leveraging raw CIR data. Various neural network architectures have been proposed, including combinations of a CNN with Long short-term memory (LSTM) or a Gate Recurrent Unit (GRU) \cite{jiang2020uwb, liu2022uwb}. To improve performance with limited data, transfer learning techniques have also been introduced \cite{fontaine2023transfer}. The latest advancements utilize transformer-based models, which process raw CIR data, sometimes supplemented with other features, for highly effective NLOS classification \cite{tomovic2022transformer, ipintf}.

\subsubsection{\textbf{Quality Metrics (TWR/TDoA)}}
Rather than using a binary LOS/NLOS classification approach, a second category of scientific publications assign a 'quality' to each link, and exclude the bad links if their configurable 'quality' metric is too low. 
However, this approach still results in a loss of information. In \cite{zhao2019select}, the anchor selection problem is formulated as an optimization problem to select the $k$ highest quality anchors. The authors of \cite{duong2024errormitigationtdoauwb} propose an unsupervised anchor selection method using deep-embedded clustering with an autoencoder to group UWB features and rank clusters based on quality. Interestingly, they show that being strict when excluding anchors quickly results in a performance decrease, potentially showing that excluding links (even bad ones) removes useful information, which further motivates our approach. 
\subsubsection{\textbf{DDoA Selection (TDoA)}}
This approach is similar to the previous ones, but instead of selecting separate links, anchor pairs (their DDoA) are filtered out, using information from both anchors. The authors of \cite{van2021anchor} propose an anchor pair selection algorithm for DDoA outlier removal using a CNN. In \cite{ma2007nonline}, a DDoA is selected using an algorithm that uses the distribution function of hyperbolic intersections of the DDoA's. If there are no NLOS errors, the intersections will be concentrated near the actual position. 
\subsection{Error Correction}
In contrast to link exclusion methods, error correction methods try to correct, rather than remove, the input features of the transmission that are used as input for the localization. 
\subsubsection{\textbf{Range Correction (TWR)}}
Several studies propose methods for UWB range error correction. Early work used feature-based machine learning, such as Support Vector Machines (SVM) and Gaussian Processes (GP), to estimate errors \cite{wymeersch2012machine}. More recent supervised approaches often use autoencoders to extract features from the CIR \cite{li2023variational, jaron}. While these methods improve ranging performance, their reliance on extensive labeled data limits their real-world applicability and generalization.
To address this, semi-supervised methods have been introduced which leverage both labeled and unlabeled data \cite{li2021semi}. Most recently, self-supervised techniques have emerged that eliminate the need for pre-labeled data entirely, for instance, by using reinforcement learning or by jointly estimating position and distance corrections \cite{coppensselfsupervised, yangselfsupervised}.
\subsubsection{\textbf{DDoA Correction (TDoA)}}
Scientific literature on TDoA error correction is far less extensive. The only existing work proposes correcting the DDoA by feeding the CIRs from anchor pairs into a CNN \cite{van2021anchor}. However, this approach has significant scalability issues, as it requires a separate forward pass for every anchor pair, leading to redundant processing of each CIR as the number of anchors grows. Moreover, the results in \cite{van2021anchor} showed no significant accuracy improvement in NLOS conditions. We compare our approach with this one in Section \ref{sec:sotacomp}.
\subsubsection{\textbf{Position Correction (TDoA)}}
Rather than correcting each anchor node pair (e.g. each DDoA separately), it is also possible to correct the position based on DDoA information from all anchor nodes within range. This much more scalable approach is proposed in this work using a transformer architecture that incorporates the CIR of all anchors to refine the estimated position. This method leverages all anchor information, even 'bad' links to detect and correct errors.
\subsection{Summary}
Table \ref{tab:relatedwork} provides a detailed comparison of all related works for TDoA improvement, highlighting the different goals, methods, and inputs used. 
In addition to providing an alternative approach, the Table also shows that our work significantly improves the scale at which a solution is evaluated, both in terms of complexity of the environment (more obstacles) as well as in scale (more anchor nodes): all works from Table \ref{tab:relatedwork} are also evaluated in relatively simple environments with only a limited number of anchors in NLOS conditions.
\subsection{Use of raw CIR in other domains}
While our work focuses on UWB, the principle of leveraging detailed channel information for positioning is well-established in other domains. Wi-Fi positioning has seen significant advances using Channel State Information (CSI), where deep learning models extract location-specific features from the frequency response of the channel \cite{zhang2021deep}. Similarly, recent research in 5G has utilized its own CIR or CSI, showing the versatility of physical layer data across different wireless technologies \cite{alawieh20235g}. Our work builds on this broader context, applying state-of-the-art sequence models to the uniquely high-resolution CIR data available from UWB systems.

\section{System and problem description}
\label{sec:problem}
This section describes the operation of TDoA localization systems. TDoA systems assume a coordinate system with a reference point (typically ref = (0, 0, 0)) to provide relative three-dimensional positions for both the fixed anchor nodes and the mobile tag. 
A total of \(N_{\text{total}}\) anchors,
$
   A = \{A_{1}, A_{2}, \ldots, A_{N_{\text{total}}}\}
$
with known positions, receive a single UWB packet transmitted by a mobile tag whose true position is denoted by $p_{\text{true}} = (p_{x},\, p_{y},\, p_{z})$
Each anchor \(A_{n}\) is located at a known coordinate $A_{n} = (A_{nx},\, A_{ny},\, A_{nz})$ and records the reception timestamp \(t_{n}\) of the packet. For simplicity, we define the true Euclidean distance between the tag's true position $p_{\text{true}}$ and an anchor \(A_{n}\) as
\begin{equation}
d_n = \sqrt{(p_{x}-A_{nx})^2 + (p_{y}-A_{ny})^2 + (p_{z}-A_{nz})^2}
\end{equation}
The true time of arrival at anchor \(A_n\) is denoted by $t_n$. An edge node computes the time differences between pairs of anchors. For any two anchors with indices \(i\) and \(j\) (where \(i, j \in \{1, 2, \ldots, N_{\text{total}}\}\) and \(i \neq j\)), the true time difference of arrival is defined as
\begin{equation}
\Delta t_{ij} = t_{i} - t_{j} = \frac{d_i - d_j}{c}
\end{equation}
Multiplying this time difference by the speed of light \(c\) yields the true DDoA:
\begin{equation}
\text{DDoA}_{ij} = c\, \Delta t_{ij} = d_i - d_j
\end{equation}
However, in NLOS conditions, multipath propagation introduces timing errors. Let the measured reception time at anchor \(A_n\) be denoted as \(t'_n\), which may differ from the true arrival time \(t_n\). The measurement error can be expressed as
\begin{equation}
\epsilon_n = t'_n - t_n
\end{equation}
The measured DDoA is therefore
\begin{equation}
\widehat{\text{DDoA}}_{ij} = c\, \Delta t'_{ij} = c\, \Delta t_{ij} + c\,(\epsilon_i - \epsilon_j)
\end{equation}
Let $\mathcal{P}$ denote the set of all index pairs corresponding to distinct anchor combinations, \(\mathcal{P} = \{(i,j) | i,j \in \{1, 2, \ldots, N_{\text{total}}\}, i \neq j\}\). For any candidate tag position \(p = (x, y, z)\), we define the distance function to anchor \(A_n\) as
\begin{equation}
d_n'(p) = \sqrt{(x-A_{nx})^2 + (y-A_{ny})^2 + (z-A_{nz})^2}
\end{equation}
A TDoA-based estimate of the tag position, denoted by \(p_{\text{TDoA}}\), is obtained by simultaneously solving the hyperboloid equations for $(x,y,z)$ for multiple \((i,j) \in \mathcal{P}\) in a non-linear least squares sense:
\begin{equation}
p_{\text{TDoA}} = \arg\min_{p} \sum_{(i,j)\in\mathcal{P}} \Big( [d_i'(p) - d_j'(p)] - \widehat{\text{DDoA}}_{ij} \Big)^2
\end{equation}
The measured DDoA equations form hyperboloids in 3D space, and the true tag position lies at the intersection of these hyperboloids in the absence of measurement errors. 

To improve the positioning accuracy in challenging environments, the CIR is leveraged to gain detailed insights into the signal propagation. Each anchor records a complex-valued CIR that captures the timing and amplitude of multiple signal paths. The received signal at anchor \(A_{n}\) is modeled as
\begin{equation}
r_{n}(t) = s(t) * \text{CIR}_{n}(t) + w_{n}(t)
\end{equation}
where \(s(t)\) is the transmitted pulse, \(*\) denotes convolution, and \(w_{n}(t)\) represents additive noise. The CIR is described by
\begin{equation}
\text{CIR}_{n}(t) = \sum_{s=1}^{S} \alpha_{s,n}\,\delta(t-\tau_{s,n}) + n_n(t)
\end{equation}
with \(S\) representing the number of multipath components, \(\alpha_{s,n}\) and \(\tau_{s,n}\) denoting the amplitude and time delay of the \(s\)th component at anchor \(n\), \(\delta(\cdot)\) the Dirac delta function, and \(n_n(t)\) the inherent channel noise at anchor \(n\). Leveraging the CIR data, our proposed error correction model is used to enhance the TDoA positioning. The model is designed to learn an optimized mapping function, denoted by \(\mathcal{F}\), that takes as input the initial TDoA-based estimated position \(p_{\text{TDoA}}\) and the set of CIR measurements from all anchors, and outputs a corrected tag position. Formally, the function is defined as
\begin{equation}
p_{\text{corr}} = \mathcal{F}\left( p_{\text{TDoA}},\, \{\text{CIR}_{n}(t)\}_{n=1}^{N_{\text{total}}} \right)
\end{equation}
The objective is to optimize \(\mathcal{F}\) such that the corrected position \(p_{\text{corr}}\) closely approximates the true tag position \(p_{\text{true}}\).
\section{Data collection and description}
\label{sec:experiment}

\begin{figure}[h]
    \centering
     \begin{subfigure}{0.48\textwidth}      \includegraphics[width=0.85\linewidth]{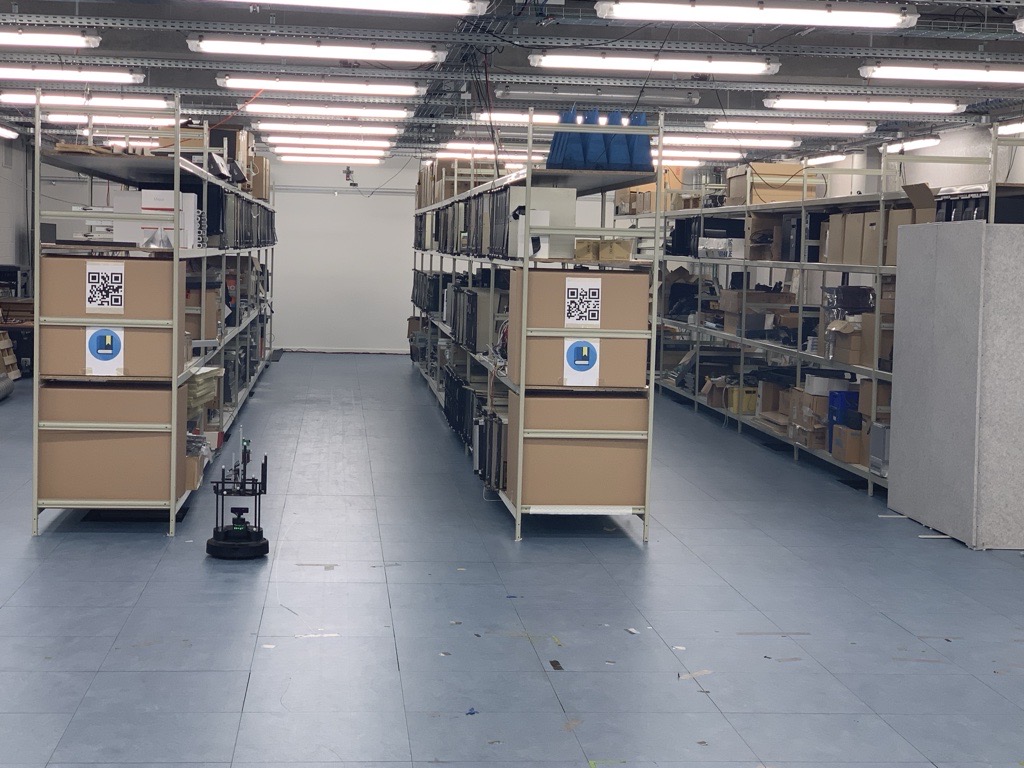}
         \caption{The IIoT lab environment}
         \label{fig:iiotpic}
     \end{subfigure}
     \begin{subfigure}{0.48\textwidth}
    \includegraphics[width=0.88\linewidth]{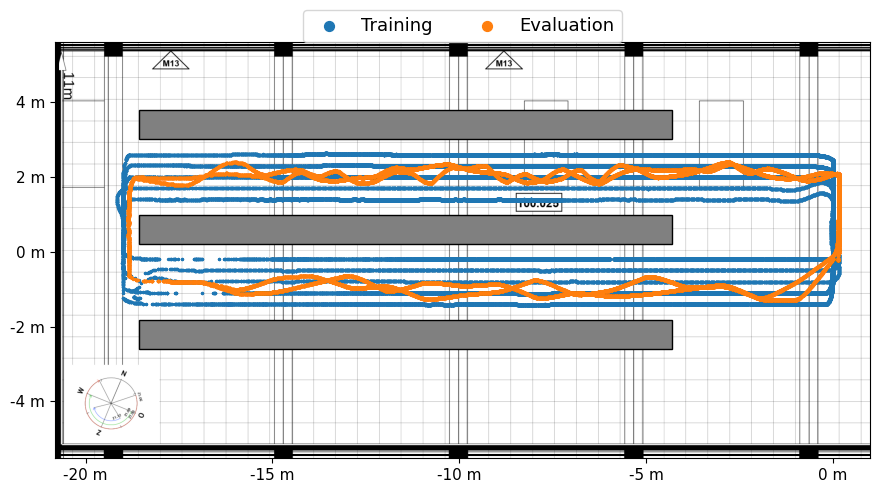}
    \caption{Floorplan of the IIoT lab with the ground truth of the training trajectory in blue and the evaluation trajectory in orange.}
    \end{subfigure}
    \caption{(a) The IIoT lab environment and (b) a comparison of the trajectories of the training datasets (blue) and evaluation dataset (orange) showing also the three racks (grey) causing severe NLOS conditions.}
    \label{fig:datasets}
\end{figure}
To evaluate the proposed method, two datasets were recorded in an industrial environment, part of the IIoT testbed \cite{iiot}, with metal racks as obstacles. Anchors were deployed in both open and NLOS locations (where several anchors are certainly obstructed). The environment spans approximately 30m × 10m. Measurements were performed using Wi-Pos devices \cite{wipos}, a platform designed for data collection with a sub-GHz wireless backbone integrated with UWB technology based on the widely used DW1000. The environment is equipped with 18 Qualisys Miqus M3 Motion Capture (MOCAP) cameras with quantified uncertainty in the millimeter range at speeds up to 340 Hz, enabling accurate ground truth determination of the position of the mobile tag. The UWB radio was configured to use channel 5 with a 64 MHz Pulse Repetition Frequency (PRF), a preamble length of 1024 symbols and a data rate of 6.8 Mbps. The tag was installed on a mobile robotic platform allowing to drive trajectories through the lab. As visualized in Figure \ref{fig:datasets}, two distinct trajectories were captured. The \textbf{training} dataset consists of systematic, straight-line paths. In contrast, the \textbf{evaluation} dataset follows a more random path. This ensures the evaluation trajectory passes through locations between the training paths. This provides a robust assessment of the model's ability to generalize and interpolate in areas not explicitly covered during training. Dataset specifics are detailed in Table \ref{tab:datasets}.
\footnote{The dataset and code used in this research are being prepared for public release and will be made available on github to ensure reproducibility. \url{https://github.com/dietercoppens/UWB_DW1000_TDoA_CIR}}
\begin{table}
    \centering
    \caption{Details of the datasets used in this research.}
    \begin{tabular}{l|ccc}
        \textbf{Name} & \textbf{ \# Anchors} & \textbf{ \# Positions}  &\begin{tabular}{@{}c@{}}\textbf{Uncorrected} \\ \textbf{TDoA MAE}\end{tabular} \\ \hline
        Training & 15 &  15505  & 1.41m\\ 
        Evaluation & 15 & 7123  & 1.48 m\\
    \end{tabular}
    \label{tab:datasets}
\end{table}
\looseness=-1
\subsection{Evaluation metrics}
For performance evaluation, we will use two metrics (1) the mean absolute error (MAE), of the Euclidean distance to the ground truth as it encapsulates the performance in a single value and (2) the circular error probability (CEP), defined as the radius of a circle, centered around the ground truth, where X\% (e.g., 95\% in the case of CEP95), of the estimated positions are within.

\section{Transformer architectural design}
\label{sec:methodology}
\begin{figure}
    \centering
    \includegraphics[width=0.75\linewidth]{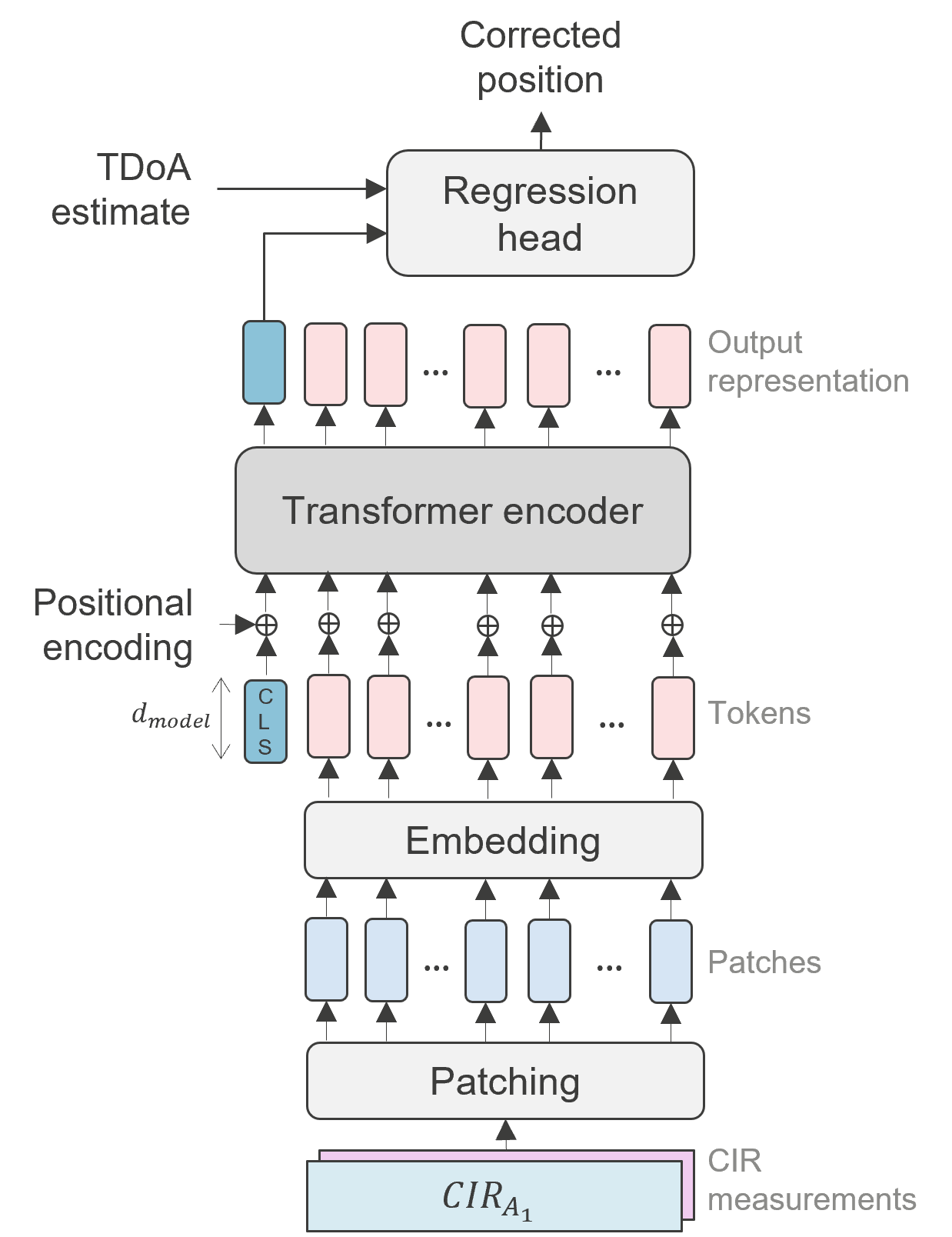}
    \caption{Schematic overview of the TDoA correction method.}
    \label{fig:tfarch}
\end{figure}
We adopted the transformer architecture for TDoA position correction due to its unique ability to model complex, non-local dependencies and its efficiency in handling inputs from multiple anchors. Unlike CNNs, which operate on fixed local receptive fields, or RNNs, which process data sequentially, the transformer's self-attention mechanism enables it to model the global context by computing pairwise interactions between all input tokens simultaneously. In our application, this allows the model to dynamically learn which parts are most relevant for correcting a position estimate, effectively suppressing noisy links while emphasizing reliable ones. This approach is also highly scalable; whereas prior CNN-based methods for TDoA often require a separate forward pass for every anchor pair, our model processes all CIRs jointly in a single pass. This improves computational efficiency and, combined with our patching and positional encoding strategies, provides the flexibility to handle the variable number of anchors common in real-world deployments.
The attention mechanism of a transformer can be defined as 

\begin{equation}
    \text{att}(Q,K,V) = \underbrace{\text{softmax}\left(\frac{QK^T}{\sqrt{h}}\right)}_aV, 
\end{equation}
\looseness=-1

\noindent where $Q$ (Query), $K$ (Key), and $V$ (Value) are linear transformations of the input. Multi-headed attention splits these components into $H$ parts, processing them independently to capture diverse representation aspects. The attention weights $a$ are the dot-product of $Q$ and $K$ followed by \textit{scaling} and \textit{softmax}. In a sense, this is a similarity that can be seen as a mask to be applied to $V$.
In this work, we only use the transformer encoder, which consists of self-attention and MLP layers with residual connections and layer normalization. Since our task is position correction, we replace the original decoder with a feed-forward network to determine a corrected 3D position.
Our transformer architecture is an encoder only, illustrated in Figure \ref{fig:tfarch}.
\subsection{Input ordering}
Each UWB packet transmitted by the tag is received by multiple anchors. Each of the receiving anchor nodes records the first path timestamp $t_n$, which is used for TDoA positioning, and the $CIR_n(t)$ which is used for error correction.  First, we process the raw $CIR_n(t)$ by converting the IQ-sampled array in a complex form to an amplitude array. Next, the CIR is min-max normalized. 
The CIR is now trimmed to 150 samples, 50 samples before and 100 samples after the first path timestamp. As each CIR sample represents about 1 ns (approximately 30 cm), this window allows us to correct errors when a multipath signal that is 15 m longer (50 samples) is mistakenly detected as the first path, and to observe multipath effects up to 30 m longer (100 samples) than the estimated first path. This range should suffice for most indoor industrial environments. 
Before these CIRs can be patched they need to be ordered logically in an input tensor $M$. Where,
\[M = 
\begin{bmatrix}
CIR_1 \\
CIR_2 \\
\vdots \\
CIR_N
\end{bmatrix}
\] with a shape of $(N, 150)$ with $N$ the total number of anchors in the environment. Ordering is thus giving each CIR a row number in this tensor. In this work, we will be comparing two ordering methods. 
\subsubsection{Fixed order}
The simplest way of ordering them is using a list that is fixed beforehand. In the fixed order approach, each anchor $A_n$ is assigned a predetermined row number. However, not every anchor node will receive a packet. As such, this fixed ordering approach requires padding the rows of non-detecting anchors with zeros, to have consistent placement of CIRs in the matrix, potentially introducing significant zero inputs in large-scale environments.
\subsubsection{Time-based order}
The time-based ordering arranges CIRs according to their received timestamps, with the earliest received CIR placed first. This method can eliminate the need for zero padding if the patching strategy allows it, leading to a matrix with fewer rows. Providing a more efficient representation of the received signals but possibly requiring different positional encoding to enable the transformer to link the CIR to the correct anchor.
\subsection{Patching}
Processing each CIR sample would be computationally too complex for a transformer architecture due to the self-attention mechanism's quadratic complexity in sequence length, leading to prohibitively high memory and computation requirements for long sequences. Patching is the process of breaking down the input tensor $M$ into discrete units. These units combine several CIR samples, reducing the sequence length. These patches of samples are then embedded to tokens to be processed by the transformer. These tokens serve as the basic building blocks for further learning.
\begin{figure*}
    \centering
    \includegraphics[width=0.65\linewidth]{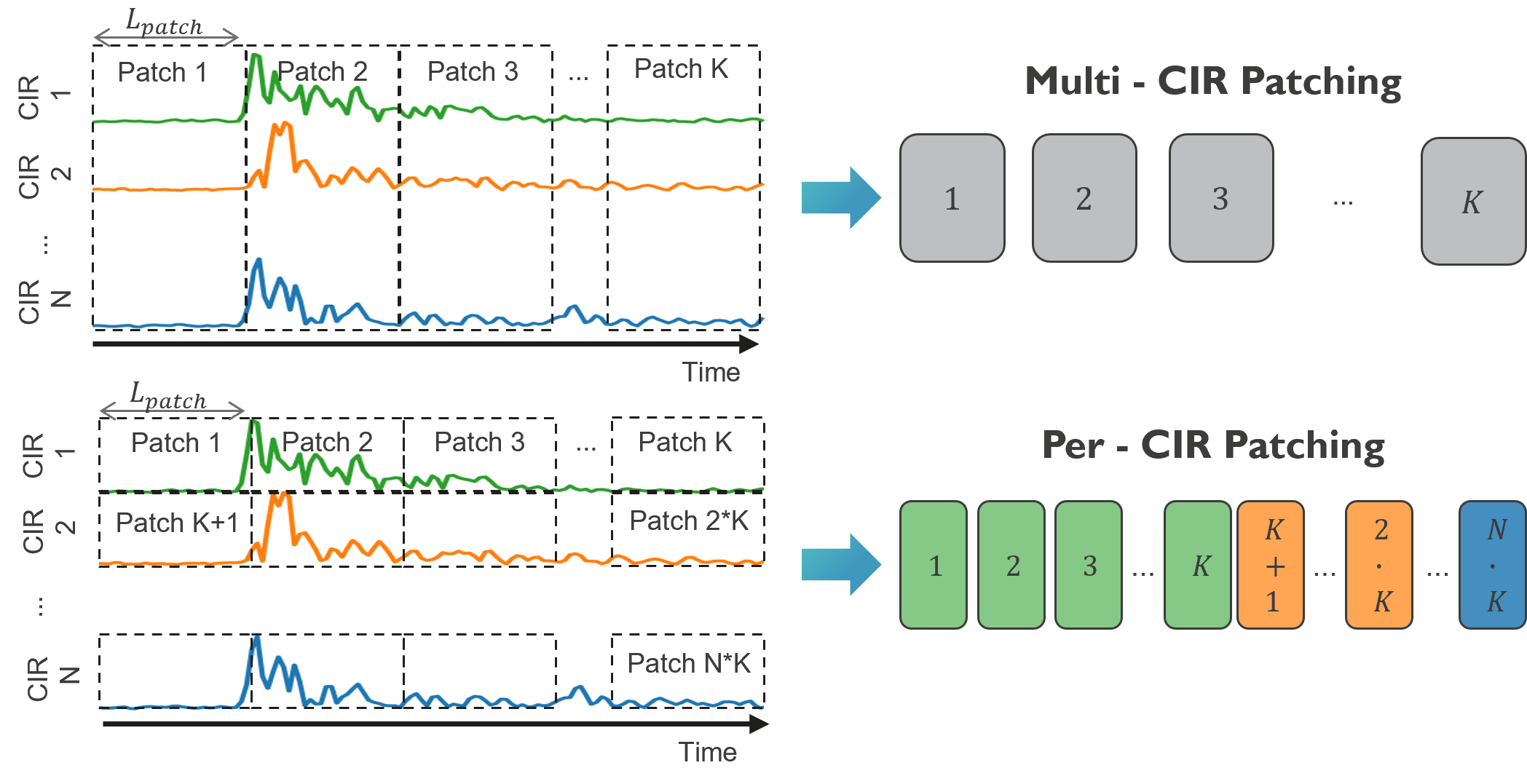}
    \caption{Illustration of the two proposed patching strategies: (top) multi-CIR patching and (bottom) per-CIR patching.}
    \label{fig:tokenization}
\end{figure*}
We propose two patching strategies illustrated in Figure \ref{fig:tokenization}, (1) multi-CIR patching and (2) per-CIR patching.
\subsubsection{Multi-CIR patching}
In multi-CIR patching, the matrix $M$ is split up into patches containing samples from all CIRs. Each patch thus has a height of $N_{total}$ and a selectable width $L_{patch}$. Leading to $K=150/L_{patch}$ multi-anchor patches 
\begin{equation}
P_k = M[:,k\cdot L_{patch}:(k+1)\cdot L_{patch}].
\end{equation}
Where $\forall k \in \{0,1,\ldots,K-1\}$
This patching is not compatible with removing the unavailable anchors as this changes the number of samples in each patch and would require different embedding networks.
\subsubsection{Per-CIR patching}
This method is an anchor-independent alternative to multi-CIR patching, meaning that each patch only contains information from a single CIR of a specific anchor. This also allows for removing the unavailable anchors instead of zero-padding, here $N$ number of available anchors for time-based ordering, compared to $N_{total}$ anchors in the environment. Each CIR is now divided into $K=150/L_{patch}$ patches leading to $N\cdot K$ anchor-independent patches 
\begin{equation}
P_{k} = M[i,\, j\cdot L_{patch} : (j+1)\cdot L_{patch}],
\end{equation}
where $\forall k \in \{0,1,\ldots,N\cdot K-1\}$, with $i = \left\lfloor \frac{k}{K} \right\rfloor$ and  $j = k \mod K$.

The motivation behind these two strategies is to explore different ways of presenting spatial and temporal information to the transformer. Multi-CIR patching forces the model to learn from the temporal alignment of signal effects across all anchors at once, treating the entire set of CIRs as a single, wide multi-channel sequence. In contrast, per-CIR patching is a more modular approach. It allows the model to first extract features from each anchor's CIR independently before using the self-attention mechanism to model the complex inter-anchor relationships. This aligns with the physical reality that each anchor provides a distinct "view" of the tag's position.

\subsection{Embedding}
The embedding converts all CIR samples in a patch into a high-dimensional latent vector or token of $d_{model}$ dimensions using a linear layer. 
\subsection{Positional encoding}
Unlike CNN's, transformers process each token independently. As a result, the inherent sequential order of the input is not captured or exploited in the architecture. To add this information positional encodings are added to insert sequential information. These encodings are added to the tokens, providing the model with explicit information about the relative or absolute position of each token in the sequence.   
\subsubsection{Learnable positional encoding}
The first method of positional encoding we use is learnable positional encoding, which introduces a trainable embedding matrix that is directly optimized during the model training process. Unlike fixed encodings, learnable positional encodings can adapt to the specific characteristics of the dataset, potentially capturing more nuanced positional relationships. This encoding adds information about where in the input sequence a token is. When the sequence order is fixed, the model can link a token using this encoding to a specific anchor and/or position in a CIR. When the order is time-based this changes, a position in the sequence is now linked to the time of arrival and not to a specific anchor.
\subsubsection{Spatial encoding}
We propose a novel positional encoding method for transformer architectures that explicitly encodes the 3D spatial coordinate of the receiving anchor. Unlike traditional positional encodings that map sequence positions, our approach transforms input coordinates (x, y, z) using sinusoidal functions with logarithmically spaced frequency bands. Let $(x,y,z)$ be the 3D coordinate of the receiving anchor. Each coordinate is normalized as follows:
$
x' = \frac{x}{X}, \quad y' = \frac{y}{Y}, \quad z' = \frac{z}{Z}.
$
With $X,Y,Z$ being the maximum possible values in the environment. We define $F$ frequency bands that are logarithmically spaced 
\begin{equation}
\omega_f = \omega_{\min}\left(\frac{\omega_{\max}}{\omega_{\min}}\right)^{\frac{f}{F-1}}, \quad f = 0, 1, \dots, F-1.
\end{equation}
The anchor positional encoding is computed by applying sinusoidal functions to each normalized coordinate
\begin{equation}
\begin{aligned}
\mathrm{PE}_{\text{spatial}}(x,y,z) = \; &\mathrm{concat}\Bigl( \bigl[\sin(x'\,\omega_f), \cos(x'\,\omega_f)\bigr]_{f=0}^{F-1},\\[1mm]
&\quad \bigl[\sin(y'\,\omega_f), \cos(y'\,\omega_f)\bigr]_{f=0}^{F-1},\\[1mm]
&\quad \bigl[\sin(z'\,\omega_f), \cos(z'\,\omega_f)\bigr]_{f=0}^{F-1} \Bigr).
\end{aligned}
\end{equation}
The concatenation in Equation (17) produces a final encoding vector of dimension $6F$, resulting from $F$ frequency bands applied to each of the 3 spatial coordinates, with 2 values (sin, cos) generated per band. To ensure this spatial encoding is compatible with the token embeddings, its dimension must match $d_{model}$. This is done by first selecting the largest F such that $6F\leq d_{model}$ and then right-padding the resulting vector with zeros. The final $PE_{spatial}$ vector can then be directly added to the token embeddings before they are fed into the transformer encoder. This encoding is not possible for multi-CIR patching as each token contains information on all anchors. For per-CIR patching with $L_{patch} < 150$ (multiple tokens from one CIR), additional positional encoding is added (relative within a CIR) that indicates the position of that token within a CIR.
The theoretical basis for introducing Spatial Encoding is to provide a strong inductive bias that is highly relevant for a localization task. Standard positional encodings only describe the order in a sequence, ignoring the physical geometry of the sensors. By explicitly encoding the anchors' 3D coordinates, we directly inform the model of the physical location from which each signal originated. This should enable the model to build a more accurate internal representation of the environment and better interpret the geometric relationships between the different CIRs, which could be helpful for correcting position errors.
\subsubsection{Spatial and Time difference encoding}
To explicitly encode the temporal separation between CIRs, we propose a time difference encoding that maps this difference using a sinusoidal function, analogous to the spatial encoding. This is then added to the spatial encoding, forming a combined representation that captures both spatial coordinates and temporal differences. This encoding is inapplicable for multi-CIR patching, as each token in that scenario already contains information from all anchors.
\subsection{General transformer architecture}
Unlike vanilla encoder-decoder transformers \cite{vaswani2017attention}, our model employs an encoder-only transformer architecture, as the fixed input and output sequence lengths eliminate the need for autoregressive generation during inference. This design enables unconstrained bidirectional information flow along the input sequence. The architectures we use consist of six of these blocks.
The output representation of the transformer encoder has the same shape as the input. To simplify the regression head we use only the CLS token output representation as input to the regression head, combined with the initial TDoA position estimate. The regression head is an MLP that determines the corrected 3D position through a series of fully connected layers with 256, 128, 64 and 3 neurons.
\section{Performance Analysis}
\label{sec:results}
This section will evaluate the performance of our proposed approaches. This section will evaluate the performance of our proposed approaches. To find the optimal setup for both patching approaches, we performed a comprehensive parameter sweep, testing \textbf{252 configurations} in total. The sweep varied key parameters: patching strategy, CIR-ordering, positional encoding, patch length ($L_{patch}$), and model dimension ($d_{model}$), as outlined in Table \ref{tab:sweep}. Note that the minimum $L_{patch}$ the per-CIR architecture was higher to avoid excessive training times. For baseline comparison, the MAE without any TDoA error correction is \textbf{1.48m}.
The results are analyzed using boxplots. Each plot visualizes the distribution of the final MAE from all experiments where one parameter was held constant. For instance, the box for a specific $d_{model}$ value summarizes the performance of all configurations run with that dimension, across every other tested parameter variation.
\begin{table*}[h]
\centering
\caption{The parameter space that was varied during the evaluation of the proposed UWB TDoA correction approach.}
\label{tab:sweep}
\begin{tabular}{ccccc}
\hline
\textbf{Patching} & \textbf{CIR-ordering} & \textbf{Positional encoding} & $\mathbf{L_{patch}}$ & $\mathbf{d_{model}}$ \\ \hline
\multirow{2}{*}{Multi-CIR patching} & Fixed & \multirow{2}{*}{Learned} & \multirow{2}{*}{1, 3, 5, 6, 10, 15, 30, 50, 75} & \multirow{2}{*}{8, 16, 32, 64, 128, 256} \\ \cline{2-2}
 & Time (zeros padded) &  &  &  \\ \hline
\multirow{6}{*}{Per-CIR patching} & \multirow{3}{*}{Fixed} & Learned & \multirow{6}{*}{6, 15, 30, 50, 75, 150} & \multirow{6}{*}{32, 64, 128, 256} \\ \cline{3-3}
 &  & Spatial &  &  \\ \cline{3-3}
 &  & Spatial + time &  &  \\ \cline{2-3}
 & \multirow{3}{*}{Time (no zeros padded)} & Learned &  &  \\ \cline{3-3}
 &  & Spatial &  &  \\ \cline{3-3}
 &  & Spatial + time &  &  \\ \hline
\end{tabular}
\end{table*}
The model architecture consists of a transformer encoder with 4 layers, each using 8 attention heads and a feed-forward network dimension of 256. Dropout regularization is applied with a probability of 0.15. The training uses the Adam optimizer, a batch size of 64, and MSE loss function. The learning rate schedule implements linear warmup for the first 5\% of training steps, reaching 0.001 followed by linear decay. Training continues for a maximum of 350 epochs with early stopping based on validation loss.

\subsection{Multi-CIR patching parameter analysis}
\begin{figure*}[ht]
    \centering
    \includegraphics[width=0.9\textwidth]{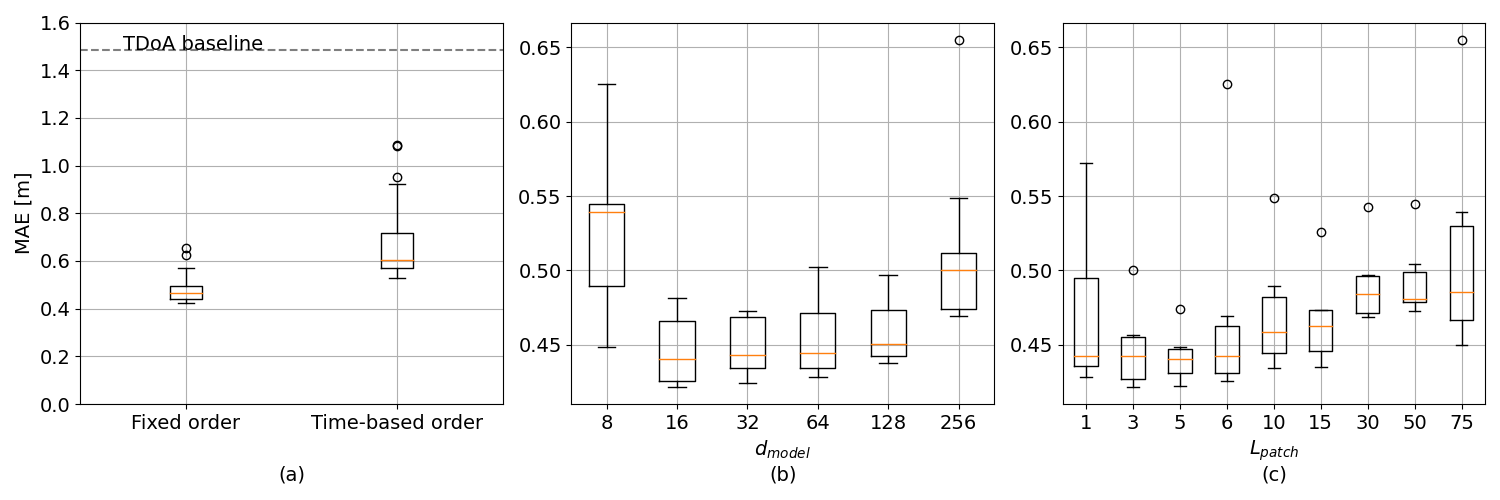}
    \caption{Results of the parameter sweep for Multi-CIR patching. Each boxplot shows the distribution of MAE across multiple experimental configurations. (a) Comparing fixed vs. time-based CIR-ordering. (b) Impact of the embedding dimension ($d_{model}$) for the fixed-order strategy. Each box aggregates results across all tested $L_{patch}$ values. (c) Impact of the patch width ($L_{patch}$) for the fixed-order strategy. Each box aggregates results across all tested $d_{model}$ values.}
    \label{fig:multioverall}
\end{figure*}
Multi-CIR patching, illustrated in Figure \ref{fig:tokenization}, divides the ordered CIR matrix $M$, where each row represents a CIR, in patches with CIR-samples from all anchors. The height of the patch is thus always the number of anchors, while the patch width $L_{patch}$ is selectable.
The evaluation results for the multi-CIR patching is summarized in Figure \ref{fig:multioverall}. 
\subsubsection{Impact of CIR-ordering}
We analyzed two CIR-ordering methods: a fixed order, where each anchor has a predetermined input position, and a time-based order, where CIRs are sorted by reception timestamp. We hypothesized that the fixed order would be superior, as it provides a consistent mapping between the input data and its source anchor. The results in Figure \ref{fig:multioverall}a confirm this expectation. For the multi-CIR strategy, fixed ordering achieved a median MAE of 0.47m, superior to the 0.60m median MAE of the time-based order. The best fixed-order configuration reached an MAE of 0.42m (a 71.6\% improvement over the baseline), while the best time-based result was 0.53m. While still a 64.2\% improvement, this is 26.2\% worse than the fixed-order optimum. The performance drop occurs because, with multi-CIR patching, the model cannot determine the origin of the CIR information when the input sequence is unpredictable. Positional encodings are added at the embedding level, leaving no subsequent mechanism to inform the model of the source anchor. We conclude that for multi-CIR patching, a fixed CIR order is critical for maximizing performance.

\subsubsection{Impact of the embedding dimension}
The embedding dimension (denoted as $d_{model}$) refers to the size of the latent vector (token) into which each patch of the CIR is projected before processing by the transformer. The expectation is that too small an embedding dimension would overly compress the rich information from the CIR, whereas excessively high dimensions might introduce noise or lead to overfitting.
Figure \ref{fig:multioverall}b illustrates the influence of the embedding dimension $d_{model}$ on the performance of the model. To improve the clarity of the figures, the results of the time-based order have been removed from the results, as the greater variance in the time-based order results made the trends less clear. For $d_{model}=8$, the median and variance of the MAE are the highest, suggesting that this dimensionality is likely too small, leading to a severe compression of information in the embedding. For the other configurations, a trend is visible where the MAE increases as $d_{model}$ becomes larger, suggesting that higher dimensionality of the tokens indeed introduce noise or reduce the efficiency of the embedding. This confirms that there is a “sweet spot” in the embedding dimension range and selecting an appropriately moderate embedding size is essential for balancing expressiveness and noise. 
\subsubsection{Impact of the patch width}
In the multi-CIR strategy, the patch width $L_{patch}$ determines how many CIR samples are grouped to form a token. Figure \ref{fig:multioverall}c illustrates its influence, showing that $L_{patch}=1$ yields the highest MAE and variance, suggesting it fails to capture necessary temporal dependencies. For other configurations, the MAE generally increases with larger $L_{patch}$ values, though an exception occurs where the best configuration at $L_{patch}=75$ outperforms those at$L_{patch}=30$ and $L_{patch}=50$. The quality of the initial token representation is critical. With very narrow patches (e.g., $L_{patch} \leq 5$), each token is a low-level signal fragment, forcing the model to first learn to assemble meaningful features (like pulse shapes) before it can model higher-level interactions. As patch width increases, each token embeds a more informative signal part, providing a better inductive bias.
However, for the multi-CIR strategy specifically, overly wide patches degrade performance. This is because a wide patch combines information from all anchors over a long duration into a single vector. This leads to excessive information compression and the loss of fine-grained signal details from individual anchors during the embedding process.

\subsection{Per-CIR patching parameter analysis}
In this Section we discuss the evaluation results of per-CIR patching. In per-CIR patching, each CIR (or row in $M$) is itself split up in patches. This is illustrated in Figure \ref{fig:tokenization}, each anchor's CIR is split into segments of size $L_{patch}$ independently, creating several (or one) patches per available anchor.
\begin{figure*}[ht]
    \centering
    \includegraphics[width=0.65\textwidth]{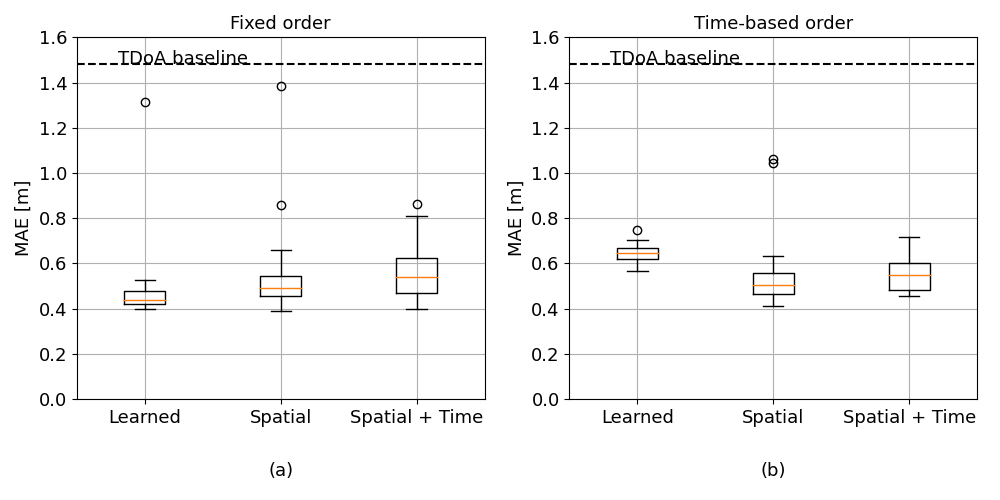}
    \caption{MAE results from the parameter sweep for Per-CIR patching, comparing different positional encoding methods. Each box aggregates results across all tested $L_{patch}$ and $d_{model}$ values for that specific encoding method. (a) Results for fixed CIR-ordering. (b) Results for time-based CIR-ordering. Highlighting that learned encodings only perform well with fixed CIR-ordering, whereas spatial encoding achieves consistent performance regardless of the CIR-ordering method.}
    \label{fig:posenc}
\end{figure*}
\subsubsection{Impact of positional encoding} 
Positional encoding adds positional information to token embeddings. For per-CIR patching, we tested learned encoding, spatial encoding (using anchor 3D coordinates), and a combined spatial/time encoding. We hypothesized that explicitly encoding anchor coordinates would help the model map tokens to their origins, particularly for time-based ordering where a fixed sequence is absent. Figure \ref{fig:posenc} shows the results for per-CIR patching. With fixed CIR-ordering (Figure \ref{fig:posenc}a), learned encoding performed most consistently (median MAE 0.48m), while spatial encoding achieved the single best result (MAE 0.39m) despite higher variability (median MAE 0.55m). The combined encoding was least consistent (median MAE 0.56m). For time-based CIR-ordering (Figure \ref{fig:posenc}b), the performance of learned encoding degraded significantly (median MAE 0.64m). In contrast, spatial encoding maintained its strong, consistent performance (median MAE 0.55m, lowest MAE 0.41m), closely matching its fixed-order results. The performance of the combined encoding also slightly diminished (lowest MAE 0.45m). This consistency demonstrates that spatial encoding successfully enables the model to relate tokens to the correct anchors without relying on a fixed input order. It is thus particularly valuable for time-ordering, as it allows for complexity reduction by removing zero padding. The combined spatial and time encoding showed no clear benefit over using spatial encoding alone.

\begin{figure*}[ht]
    \centering
    \includegraphics[width=0.65\textwidth]{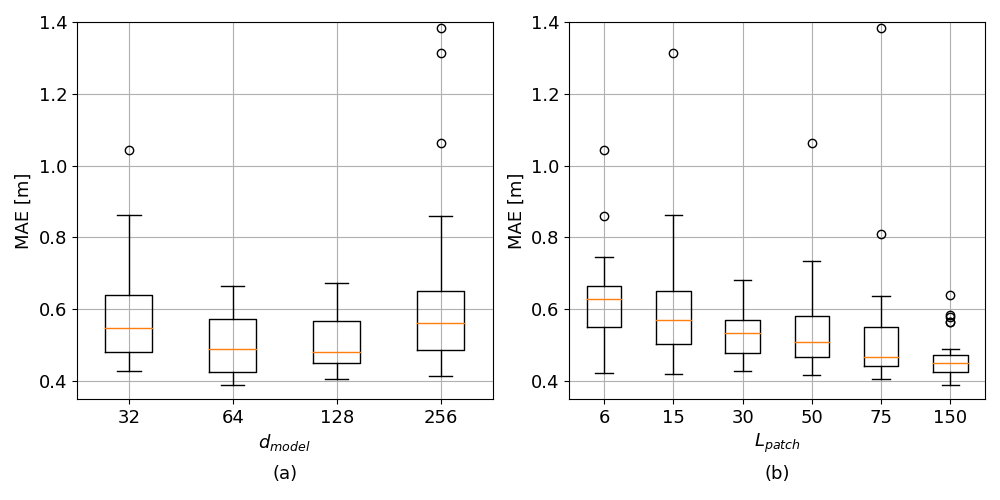}
    \caption{Results of the parameter sweep for Per-CIR patching, analyzing the impact of model dimension and patch width. Each box aggregates results across all tested $L_{patch}$ and $d_{model}$ values. (a) Impact of the embedding dimension ($d_{model}$). (b) Impact of the patch width ($L_{patch}$).}
    \label{fig:percirdpatch}
\end{figure*}
\subsubsection{Impact of embedding size} 
The embedding dimension, $d_{model}$, is the size of the latent vector representing each CIR patch. A small dimension may fail to capture detailed features, while a large one risks overfitting. As illustrated in Figure \ref{fig:percirdpatch}a, the results show optimal performance at $d_{model}$ values of 64 and 128, which yield median MAEs around 0.48 - 0.50m and exhibit more consistent performance, indicated by their compact interquartile ranges. In contrast, both smaller (32) and larger (256) dimensions result in slightly worse median MAEs around 0.55m with higher variance. Notably, $d_{model}=256$ produces significant outliers up to 1.4m. This indicates that a moderate embedding size is crucial for robust feature extraction in per-CIR patching, as increasing beyond 128 leads to degraded and less stable results.
\subsubsection{Impact of patch width}
In the per-CIR strategy, the patch width $L_{patch}$ defines how many consecutive CIR samples from an anchor are grouped into one token. As shown in Figure \ref{fig:percirdpatch}b, performance consistently improves with increasing patch size, peaking at $L_{patch}=150$. Here, wider patches are beneficial because each token becomes a more complete representation of a single anchor's CIR. The superior performance at $L_{patch}=150$ is due to a key architectural implication: at this size, each anchor's entire 150-sample CIR is embedded into a single token. This forces the model into a hierarchical two-stage process:(1) he embedding layer learns to create a single, holistic vector summarizing the entire channel response for each anchor.  (2) The self-attention mechanism then operates at a higher level of abstraction, modeling the relationships between these N anchor-level summary tokens. The strong results of this configuration suggest that learning a complete representation of each CIR before modeling the inter-anchor dependencies is a highly effective strategy for TDoA correction. In contrast, when $L_{patch}=75$, each CIR is split into two tokens. This requires the attention mechanism to operate on a more complex sequence of 2N tokens representing only partial CIR information, which our results show is less effective.
\section{Complexity Analysis}
\label{sec:complexity}
The computational complexity of our approach is mainly determined by the patching strategy. The CIR-ordering or more specifically, the possibility of eliminating zero-padding is the second largest effect. The first part of the model is the embedding of the input sample patches to a token. This is a linear layer that maps $k_{samples}$ to $d_{model}$ embedding dimensions, for $n$ tokens, giving a complexity of $\mathcal{O}(k_{samples} \cdot d_{model} \cdot n)$. Second is the self-attention mechanism that computes attention scores for all token pairs, with complexity $\mathcal{O}(n^2 \cdot d_{model})$. Finally, The feed-forward network applies two linear transformations to each token, resulting in a complexity of $\mathcal{O}(n \cdot d_{model} \cdot d_{ff})$. With $d_{ff}$ the dimension of the intermediate feed-forward layer. To calculate the total number of operations, the self-attention and feed-forward complexity need to be multiplied by the number of encoder layers (4 in our approach) and the complexity of the final regression head needs to be added as well.
\begin{table*}[h]
\centering
\caption{Complexity of the embedding, self-attention and feed-forward layers for different patching and CIR-ordering approaches, $N_{total}$ is the total number of anchors in the environment, $N_{av}$ the available anchors at a position, $L_{patch}$ the patch width, $d_{model}$ the embedding dimension and $d_{ff}$ the feed-forward dimension}
\label{tab:complexity}
\begin{tabular}{llll}
\hline
 & \multicolumn{1}{c}{\textbf{Embedding}} & \multicolumn{1}{c}{\textbf{Self-attention}} & \multicolumn{1}{c}{\textbf{Feed-Forward}} \\ \hline
\textbf{Multi-CIR} & $\mathcal{O}(N_{total} \cdot 150 \cdot d_{model})$ & $\mathcal{O}((150/L_{patch})^2 \cdot d_{model})$ & $\mathcal{O}((150/L_{patch})\cdot d_{model} \cdot d_{ff})$ \\ 
\textbf{Per-CIR fixed} & $\mathcal{O}(N_{total}  \cdot 150 \cdot d_{model})$ & $\mathcal{O}((N_{total} \cdot(150/L_{patch}))^2 \cdot d_{model})$ & $\mathcal{O}(N_{total} \cdot(150/L_{patch}) \cdot d_{model}\cdot d_{ff})$ \\ 
\textbf{Per-CIR time} & $\mathcal{O}(N_{av} \cdot 150 \cdot d_{model})$ & $\mathcal{O}((N_{av} \cdot (150/L_{patch}))^2 \cdot d_{model})$ & $\mathcal{O}(N_{av} \cdot(150/L_{patch}) \cdot d_{model}\cdot d_{ff})$ \\ \hline
\end{tabular}
\end{table*}

For the multi-CIR patching, $n = 150/L_{patch}$ and $k_{samples} = N_{total} \cdot L_{patch}$ with $N_{total}$ the total number of anchors in the environment. For per-CIR patching this changes to $n = N \cdot (150/L_{patch})$ and $k_{samples} = L_{patch}$. However, important to note is $N$ can be variable for the per-CIR patching. For fixed CIR-ordering this value is always the number of anchors in the environment, 15 in our dataset. For time-based ordering, this is the number of available CIRs $N_{av}$ at that position. In our measured datasets, the average number of available CIRs is 6.2. For further calculations we will use this number as $N_{av}$. The different complexities for each subpart and each architectural difference are given in Table \ref{tab:complexity}.
\begin{figure*}
    \centering
\includegraphics[width=0.82\textwidth]{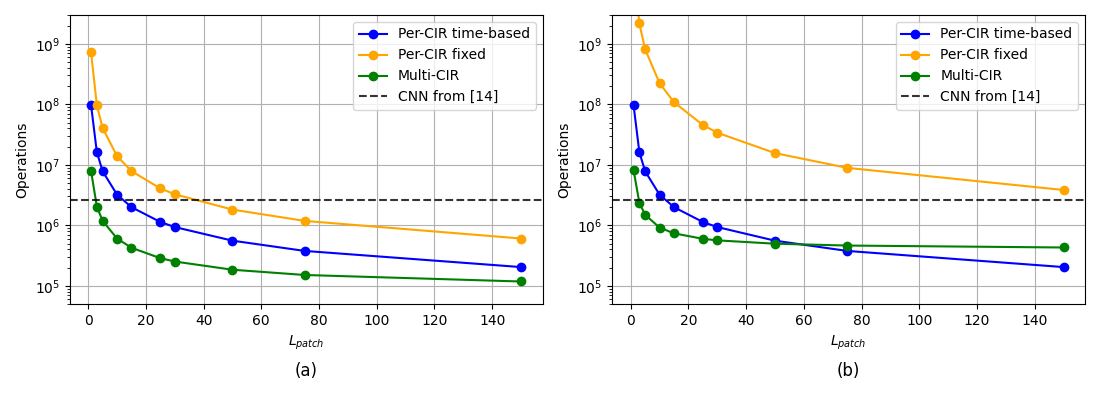}
    \caption{The total complexity as a function of the patch width (with $d_{model}=32$) for multi-CIR patching, per-CIR patching with fixed ordering, and per-CIR patching with time-based ordering and the DDoA correction approach with a CNN from \cite{van2021anchor}. (a) is in the IIoT lab with 15 anchors and 6 available on average, (b) is a theoretical large-scale environment with 50 anchors and 6 available on average.}
    \label{fig:complexfull}
\end{figure*}
The relation between $L_{patch}$ and the complexity of the architectures, for the same $d_{model}$ is illustrated in Figure \ref{fig:complexfull}a. The figure shows that the per-CIR fixed architecture exhibits the highest complexity, followed by the per-CIR time-based architecture, with the multi-CIR approach having the lowest complexity. The complexity of the multi-CIR patching decreases most quickly for increasing $L_{patch}$ because the self-attention and feed-forward complexity are not dependent on $N$, and thus lower. As $L_{patch}$ increases the gap between multi-CIR patching and per-CIR patching time-based diminishes until they both reach a similar lower bound.
\subsection{Complexity-based performance comparison}
\begin{figure}[h]    \centering    \includegraphics[width=0.85\linewidth]{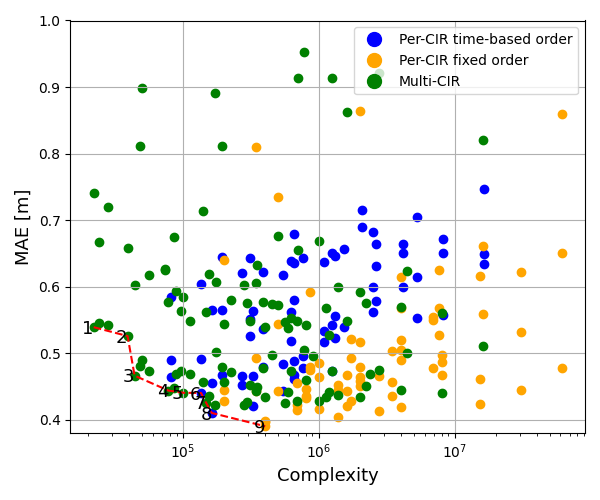}
    \caption{Comparison of computational complexity versus MAE for different configurations of per-CIR and multi-CIR approaches. The per-CIR time-based order (blue), per-CIR fixed order (orange), and multi-CIR (green) configurations are plotted. The Pareto-optimal solutions are numbered along the red line.}
    \label{fig:pareto}
\end{figure}

In Table \ref{sec:complexity}, the computational complexity of the different approaches was compared. This showed that the complexity was highly dependent on the selected parameters. Based on the complexity and the results of configurations, we can determine the Pareto-optimal solutions, highlighting the trade-offs between performance and computational cost of the different approaches, as illustrated in Figure \ref{fig:pareto}.
The Pareto-optimal solutions, indicated by the red line and detailed in Table \ref{tab:combined_pareto}, show the configurations that have the best trade-off between accuracy and complexity. Multi-CIR patching dominates the low-complexity region of the Pareto front because it delivers high performance for high $L_{patch}$ and low $d_{model}$ configurations (low complexity). For increasing complexity and performance, the per-CIR patching becomes the optimal configuration with the lowest MAE for per-CIR patching with fixed ordering.
\begin{table*}
    \centering
    \caption{Summary of Pareto-optimal configurations and their corresponding performance metrics. The first six columns describe the architectural parameters, while the remaining columns show CEP values at 50\%, 75\%, 90\%, 95\%, and 99\%.}
    \label{tab:combined_pareto}
    \begin{tabular}{ccccccccccc}
        \hline
        \textbf{Config} & \textbf{Patching} & \textbf{CIR-ordering} & $\mathbf{d_{model}}$ & $\mathbf{L_{patch}}$ & \textbf{Positional Encoding} & \textbf{CEP50} & \textbf{CEP75} & \textbf{CEP90} & \textbf{CEP95} & \textbf{CEP99} \\ \hline
        TDoA & - & - & - & - & - & 0.59 m & 1.72 m & 4.43 m & 5.37 m & 9.22 m \\ 
        1 & Multi-CIR & Fixed & 8 & 75 & Learned & 0.23 m & 0.66 m & 1.36 m & 2.03 m & 3.68 m \\ 
        2 & Multi-CIR & Fixed & 8 & 15 & Learned & 0.24 m & 0.71 m & 1.31 m & 1.81 m & 3.066 m \\ 
        3 & Multi-CIR & Fixed & 16 & 75 & Learned & 0.22 m & 0.57 m & 1.19 m & 1.66 m & 3.142 m \\ 
        4 & Multi-CIR & Fixed & 16 & 15 & Learned & 0.21 m & 0.57 m & 1.01 m & 1.54 m & 2.93 m \\ 
        5 & Multi-CIR & Fixed & 16 & 10 & Learned & 0.22 m & 0.57 m & 1.07 m & 1.48 m & 2.97 m \\ 
        6 & Per-CIR & Time-based & 32 & 150 & Spatial & 0.22 m & 0.65 m & 1.26 m & 1.67 m & 2.85 m \\ 
        7 & Multi-CIR & Fixed & 16 & 6 & Learned & 0.22 m & 0.53 m & 1.04 m & 1.47 m & 2.73 m \\ 
        8 & Per-CIR & Time-based & 64 & 150 & Spatial & 0.20 m & 0.51 m & 1.01 m & 1.42 m & 2.60 m \\ 
        9 & Per-CIR & Fixed & 64 & 150 & Spatial & 0.19 m & 0.52 m & 1.07 m & 1.34 m & 2.35 m \\ 
        \hline
    \end{tabular}
\end{table*}

All configurations significantly outperform the baseline TDoA, with the best model achieving a CEP50 of 0.19m and a CEP99 of 2.35m. The per-CIR patching method generally yields better results, especially at higher confidence levels, indicating it is more effective at mitigating large outlier errors.
This reveals a clear trade-off: multi-CIR patching is less computationally demanding and suitable for resource-constrained applications, while per-CIR patching offers superior accuracy at a higher computational cost.
Crucially, the complexity of the per-CIR approach with time-based ordering remains constant as the environment scales. In contrast, the complexity of multi-CIR and fixed-order per-CIR methods increases significantly with more anchors. Therefore, for larger deployments, the time-based per-CIR approach is expected to offer the best balance of performance and efficiency and dominate the Pareto front.

\subsection{Practical complexity metrics}
To complement our theoretical complexity analysis, the practical inference performance of our key models were measured. The latency or inference time, the model size on disk and the peak GPU memory was benchmarked on a GPU (NVIDIA GeForce GTX 1080 Ti). The practical benchmarks in Table \ref{tab:measuredperf} validate our theoretical analysis. Inference time increases from 5.45 ms to just 12.72 ms for the most complex model, a modest scaling explained by the GPU's parallel architecture achieving higher throughput on larger workloads. Since peak GPU memory usage is stable and low (~10-11 MB), all models are confirmed to be highly efficient and suitable for real-time deployment.

\begin{table}[h]
\centering
\caption{Practical complexity of Pareto-optimal configurations from Table \ref{tab:combined_pareto} with benchmarks run on a single NVIDIA GeForce GTX 1080 Ti GPU}
\label{tab:measuredperf}
\begin{tabular}{cccc}
\hline
\textbf{Config} & \textbf{\begin{tabular}[c]{@{}c@{}}Inference \\ time (ms)\end{tabular}} & \textbf{\begin{tabular}[c]{@{}c@{}}Model size \\ on disk (MB)\end{tabular}} & \textbf{\begin{tabular}[c]{@{}c@{}}Peak GPU \\ Memory (MB)\end{tabular}} \\ \hline
1 & 5.45 & 0.31 & 9.97 \\
2 & 6.02 & 0.28 & 10.25 \\
3 & 5.99 & 0.43 & 10.4 \\
4 & 6.84 & 0.31 & 10.27 \\
5 & 7.40 & 0.37 & 10.35 \\
6 & 8.26 & 0.37 & 10.36 \\
7 & 8.27 & 0.58 & 10.56 \\
8 & 9.09 & 0.37 & 10.36 \\
9 & 12.72 & 1.1 & 11.09 \\ \hline
\end{tabular}
\end{table}
\subsection{Influence of environment size}
Important to note is that the gap between fixed ordering (both per-CIR and multi-CIR) and time-based ordering will widen the larger and more complex an environment becomes. As the size of the environment increases, the total number of anchors increases but the number of anchors within reach in a certain position will remain similar. For example, an industrial warehouse can have 50 anchors instead of the 15 in our dataset. This results in a significant increase in complexity for the fixed ordering approach, requiring 50 rows in the input instead of 15, while the number of available anchors in each position will remain similar. Figure \ref{fig:complexfull}b illustrates this effect. The complexity of the per-CIR time-based ordering approach scales to more and more complex environments.

\section{Comparison with related work}
\label{sec:sotacomp}
In the following sections, our approach is compared with \cite{duong2024errormitigationtdoauwb}, an unsupervised link exclusion method using quality metrics and \cite{van2021anchor}, a DDoA correction approach. The datasets used in \cite{duong2024errormitigationtdoauwb} and \cite{van2021anchor} were recorded in the same environment as ours, however there are some differences. The authors of \cite{van2021anchor} only recorded in the open space part of the environment (no influence of the racks) but they did add some obstacles in their 'NLOS tag + anchor' situation.  The authors of \cite{duong2024errormitigationtdoauwb} excluded all positions with x-values between -21 m and -15 m due to poor signal quality, not allowing their link exclusion method to work. This is because large areas in the environment have limited LOS links where bad link exclusion has difficulty improving the positioning. For positions in the x-direction lower than -15 m (see Figure \ref{fig:datasets}), using the NLOS estimation from the Qorvo DW1000 user manual \cite{dw1000manual}, 33 \% have no LOS links, 65\% have less than two LOS links, 87\% have less than three LOS links. This shows that if this metric was used, 87\% of the positions could not be determined as there is a minimum of three links required for determining a TDoA position.

\subsection{Performance comparison}
 To ensure a fair and robust comparison against the baseline methods, we establish a common ground based on the experimental environment and baseline performance. All experiments (ours, as well as those in \cite{duong2024errormitigationtdoauwb} and \cite{van2021anchor}) were conducted in the exact same IIoT lab environment, ensuring identical physical conditions and NLoS challenges. A strong indicator of a fair comparison is the uncorrected TDoA performance, which is nearly identical across the different datasets. If we exclude the difficult region (all samples with x-values between -21 m and 14 m) from our dataset in the exact same way as done in \cite{duong2024errormitigationtdoauwb}. The baseline MAE on our dataset is 0.45 m , which perfectly matches the 0.45 m reported in \cite{duong2024errormitigationtdoauwb} and is very close to the 0.46 m in \cite{van2021anchor}. This confirms that all datasets represent a similar level of initial difficulty and NLoS impact, allowing for a meaningful comparison of the improvement each method provides. To indicate that the data splits are not identical an asterisk was added to the two baselines.

The results are summarized in Table \ref{tab:paretoresults}, which compares the performance of the baseline TDoA method, the DDoA correction approach (called 'ML error mitigation' in \cite{van2021anchor}), the DEC+k-means approach from \cite{duong2024errormitigationtdoauwb}, and the best performing/most complex configuration from the Pareto front (number 9 in Table \ref{tab:combined_pareto}): per-CIR patching with fixed ordering $d_{model}=64$, $L_{patch} = 150$ with spatial encoding.

\begin{table}[h!]
    \centering
    \begin{tabular}{ccccc}
        \hline
        \textbf{Configuration} & \textbf{MAE} & \textbf{CEP75} & \textbf{CEP90} & \textbf{CEP95} \\
        \hline
        Traditional TDoA & 0.45 m & 0.46 m & 1.04 m & 1.47 m \\
        DDoA Correction* \cite{van2021anchor} & 0.47 m & 0.56 m & / & 1.22 m \\
        DEC+k-means* \cite{duong2024errormitigationtdoauwb} & 0.33 m & 0.34 m & 0.55 m & 0.74 m \\
        \textbf{Ours (per-CIR fixed)} & \textbf{0.19 m} & \textbf{0.20 m} & \textbf{0.33 m} & \textbf{0.55 m} \\
        \hline
    \end{tabular}
    \caption{Comparison of localization performance between the traditional TDoA method, the DDoA correction approach \cite{van2021anchor}, the unsupervised DEC+k-means approach from \cite{duong2024errormitigationtdoauwb}, and our approach.}
    \label{tab:paretoresults}
\end{table}

As shown in Table \ref{tab:paretoresults}, the DDoA correction approach fails to improve the TDoA positioning performance and our approach significantly outperforms the unsupervised DEC+k-means method across all metrics. Our approach achieves a minimum MAE of 0.19 m compared to 0.33 m for DEC+k-means. Similarly, our CEP75, CEP90, and CEP95 values demonstrate substantial improvements, reflecting our method's enhanced accuracy and reliability. These results confirm that our direct position approach is an improvement compared to other approaches by consistently outperforming the (unsupervised) state-of-the-art while also enabling position corrections in regions with inferior signal quality.
\subsection{Complexity comparison}
The practical results in Table \ref{tab:measuredperf} show that our proposed models are highly efficient. Even our most accurate and complex model (Config 9) achieves a low inference latency of 12.72 ms, confirming its suitability for real-time systems.

Unfortunately, there is not enough detail about the architecture in \cite{duong2024errormitigationtdoauwb} to determine the complexity of this approach and there is thus no comparison possible. The complexity of our approach can be compared with the CNN used for DDoA correction in \cite{van2021anchor}, where each available DDoA is corrected. In our dataset, there are on average 6 CIRs available. Leading to an average of 15 available DDoAs to be corrected for one position. We can determine the complexity of the CNN by using the complexity of the convolutional layers and feed-forward layers. For each forward-pass, 173704 computations are required and this process needs to be repeated for every available pair, resulting in a total of $2.6*10^6$ operations. This is illustrated in Figure \ref{fig:complexfull} for comparison with our approach.
For most values of $L_{patch}$, both multi-CIR and per-CIR patching (with time ordering) require fewer operations than the CNN approach from \cite{van2021anchor}, showing that our approach is also computationally efficient.

\section{Conclusion}
\label{sec:conclusion}
This paper introduced a transformer-based model that directly corrects UWB TDoA position estimates using raw CIR data from all available anchors. Unlike previous methods that rely on excluding 'bad' links, our approach corrects errors, proving effective even when all links have poor signal quality. We introduced novel patching and spatial encoding strategies that ensure scalability to large, complex environments. Real-world experiments show our method improves accuracy by up to 73.6\% over the baseline TDoA and 42.4\% over a state-of-the-art unsupervised link exclusion method, enabling robust and scalable TDoA-based positioning without discarding valuable data.

%

\appendices


%
\bibliographystyle{IEEEtran}
\bibliography{sample-base}

\begin{thebibliography}{10}
\providecommand{\url}[1]{#1}
\csname url@samestyle\endcsname
\providecommand{\newblock}{\relax}
\providecommand{\bibinfo}[2]{#2}
\providecommand{\BIBentrySTDinterwordspacing}{\spaceskip=0pt\relax}
\providecommand{\BIBentryALTinterwordstretchfactor}{4}
\providecommand{\BIBentryALTinterwordspacing}{\spaceskip=\fontdimen2\font plus
\BIBentryALTinterwordstretchfactor\fontdimen3\font minus \fontdimen4\font\relax}
\providecommand{\BIBforeignlanguage}[2]{{%
\expandafter\ifx\csname l@#1\endcsname\relax
\typeout{** WARNING: IEEEtran.bst: No hyphenation pattern has been}%
\typeout{** loaded for the language `#1'. Using the pattern for}%
\typeout{** the default language instead.}%
\else
\language=\csname l@#1\endcsname
\fi
#2}}
\providecommand{\BIBdecl}{\relax}
\BIBdecl

\bibitem{surveyindoorlocapplications}
C.~Laoudias, A.~Moreira, S.~Kim, S.~Lee, L.~Wirola, and C.~Fischione, ``A survey of enabling technologies for network localization, tracking, and navigation,'' \emph{IEEE Communications Surveys \& Tutorials}, vol.~20, no.~4, pp. 3607--3644, 2018.

\bibitem{smartlogistics}
M.~Elsanhoury, P.~Mäkelä, J.~Koljonen, P.~Välisuo, A.~Shamsuzzoha, T.~Mantere, M.~Elmusrati, and H.~Kuusniemi, ``Precision positioning for smart logistics using ultra-wideband technology-based indoor navigation: A review,'' \emph{IEEE Access}, vol.~10, pp. 44\,413--44\,445, 2022.

\bibitem{coppens2022overview}
D.~Coppens, A.~Shahid, S.~Lemey, B.~Van~Herbruggen, C.~Marshall, and E.~De~Poorter, ``An overview of uwb standards and organizations (ieee 802.15.4, fira, apple): Interoperability aspects and future research directions,'' \emph{IEEE Access}, vol.~10, pp. 70\,219--70\,241, 2022.

\bibitem{20}
L.~Flueratoru, S.~Wehrli, M.~Magno, E.~S. Lohan, and D.~Niculescu, ``High-accuracy ranging and localization with ultrawideband communications for energy-constrained devices,'' \emph{IEEE Internet of Things Journal}, vol.~9, no.~10, pp. 7463--7480, 2022.

\bibitem{mao2018probabilistic}
C.~Mao, K.~Lin, T.~Yu, and Y.~Shen, ``A probabilistic learning approach to uwb ranging error mitigation,'' in \emph{2018 IEEE Global Communications Conference (GLOBECOM)}.\hskip 1em plus 0.5em minus 0.4em\relax IEEE, 2018, pp. 1--6.

\bibitem{jaron}
J.~Fontaine, M.~Ridolfi, B.~Van~Herbruggen, A.~Shahid, and E.~De~Poorter, ``Edge inference for uwb ranging error correction using autoencoders,'' \emph{IEEE Access}, vol.~8, pp. 139\,143--139\,155, 2020.

\bibitem{li2023variational}
Y.~Li, S.~Mazuelas, and Y.~Shen, ``A variational learning approach for concurrent distance estimation and environmental identification,'' \emph{IEEE Transactions on Wireless Communications}, 2023.

\bibitem{coppensselfsupervised}
D.~Coppens, B.~van Herbruggen, A.~Shahid, and E.~de~Poorter, ``Removing the need for ground truth uwb data collection: Self-supervised ranging error correction using deep reinforcement learning,'' \emph{IEEE Transactions on Machine Learning in Communications and Networking}, vol.~2, pp. 1615--1627, 2024.

\bibitem{duong2024errormitigationtdoauwb}
\BIBentryALTinterwordspacing
P.~B. Duong, B.~V. Herbruggen, A.~Broering, A.~Shahid, and E.~D. Poorter, ``Error mitigation for tdoa uwb indoor localization using unsupervised machine learning,'' 2024. [Online]. Available: \url{https://arxiv.org/abs/2404.06824}
\BIBentrySTDinterwordspacing

\bibitem{zhao2019select}
Y.~Zhao, Z.~Li, B.~Hao, P.~Wan, and L.~Wang, ``How to select the best sensors for tdoa and tdoa/aoa localization?'' \emph{China Communications}, vol.~16, no.~2, pp. 134--145, 2019.

\bibitem{ma2007nonline}
C.~Ma, R.~Klukas, and G.~Lachapelle, ``A nonline-of-sight error-mitigation method for toa measurements,'' \emph{IEEE Transactions on Vehicular Technology}, vol.~56, no.~2, pp. 641--651, 2007.

\bibitem{van2021anchor}
B.~Van~Herbruggen, J.~Fontaine, and E.~De~Poorter, ``Anchor pair selection for error correction in time difference of arrival (tdoa) ultra wideband (uwb) positioning systems,'' in \emph{2021 International Conference on Indoor Positioning and Indoor Navigation (IPIN)}.\hskip 1em plus 0.5em minus 0.4em\relax IEEE, 2021, pp. 1--8.

\bibitem{che}
F.~Che, Q.~Z. Ahmed, F.~A. Khan, and P.~I. Lazaridis, ``Anomaly detection based on generalized gaussian distribution approach for ultra-wideband (uwb) indoor positioning system,'' in \emph{2021 26th International Conference on Automation and Computing (ICAC)}, 2021, pp. 1--5.

\bibitem{fan2019non}
J.~Fan and A.~S. Awan, ``Non-line-of-sight identification based on unsupervised machine learning in ultra wideband systems,'' \emph{IEEE Access}, vol.~7, pp. 32\,464--32\,471, 2019.

\bibitem{sangnlos}
\BIBentryALTinterwordspacing
C.~L. Sang, B.~Steinhagen, J.~D. Homburg, M.~Adams, M.~Hesse, and U.~Rückert, ``Identification of nlos and multi-path conditions in uwb localization using machine learning methods,'' \emph{Applied Sciences}, vol.~10, no.~11, 2020. [Online]. Available: \url{https://www.mdpi.com/2076-3417/10/11/3980}
\BIBentrySTDinterwordspacing

\bibitem{jiang2020uwb}
C.~Jiang, J.~Shen, S.~Chen, Y.~Chen, D.~Liu, and Y.~Bo, ``Uwb nlos/los classification using deep learning method,'' \emph{IEEE Communications Letters}, vol.~24, no.~10, pp. 2226--2230, 2020.

\bibitem{liu2022uwb}
Q.~Liu, Z.~Yin, Y.~Zhao, Z.~Wu, and M.~Wu, ``Uwb los/nlos identification in multiple indoor environments using deep learning methods,'' \emph{Physical Communication}, vol.~52, p. 101695, 2022.

\bibitem{fontaine2023transfer}
J.~Fontaine, F.~Che, A.~Shahid, B.~Van~Herbruggen, Q.~Z. Ahmed, W.~B. Abbas, and E.~De~Poorter, ``Transfer learning for uwb error correction and (n) los classification in multiple environments,'' \emph{IEEE Internet of Things Journal}, 2023.

\bibitem{tomovic2022transformer}
S.~Tomovi{\'c}, K.~Bregar, T.~Javornik, and I.~Radusinovi{\'c}, ``Transformer-based nlos detection in uwb localization systems,'' in \emph{2022 30th Telecommunications Forum (TELFOR)}.\hskip 1em plus 0.5em minus 0.4em\relax IEEE, 2022, pp. 1--4.

\bibitem{ipintf}
H.~Yang, Y.~Wang, C.~Seow, M.~Sun, and D.~Plets, ``Uwb nlos identification and mitigation based on bidirectional encoder representations from transformer (bert) deep learning,'' in \emph{2024 14th International Conference on Indoor Positioning and Indoor Navigation (IPIN)}, 2024, pp. 1--6.

\bibitem{wymeersch2012machine}
H.~Wymeersch, S.~Maran{\`o}, W.~M. Gifford, and M.~Z. Win, ``A machine learning approach to ranging error mitigation for uwb localization,'' \emph{IEEE transactions on communications}, vol.~60, no.~6, pp. 1719--1728, 2012.

\bibitem{li2021semi}
Y.~Li, S.~Mazuelas, and Y.~Shen, ``A semi-supervised learning approach for ranging error mitigation based on uwb waveform,'' in \emph{MILCOM 2021-2021 IEEE Military Communications Conference (MILCOM)}.\hskip 1em plus 0.5em minus 0.4em\relax IEEE, 2021, pp. 533--537.

\bibitem{yangselfsupervised}
B.~Yang, J.~Li, Z.~Shao, and H.~Zhang, ``Self-supervised deep location and ranging error correction for uwb localization,'' \emph{IEEE Sensors Journal}, vol.~23, no.~9, pp. 9549--9559, 2023.

\bibitem{zhang2021deep}
Z.~Zhang, M.~Lee, and S.~Choi, ``Deep-learning-based wi-fi indoor positioning system using continuous csi of trajectories,'' \emph{Sensors}, vol.~21, no.~17, p. 5776, 2021.

\bibitem{alawieh20235g}
M.~Alawieh and G.~Kontes, ``5g positioning advancements with ai/ml,'' \emph{arXiv preprint arXiv:2401.02427}, 2023.

\bibitem{iiot}
\BIBentryALTinterwordspacing
 [Online]. Available: \url{https://idlab.ugent.be/resources/industrial-iot-lab}
\BIBentrySTDinterwordspacing

\bibitem{wipos}
B.~Van~Herbruggen, B.~Jooris, J.~Rossey \emph{et~al.}, ``Wi-pos: A low-cost, open source uwb hardware platform with long range sub-ghz backbone,'' \emph{Sensors}, vol.~19, no.~7, p. 1548, 2019.

\bibitem{vaswani2017attention}
\BIBentryALTinterwordspacing
A.~Vaswani, N.~Shazeer, N.~Parmar, J.~Uszkoreit, L.~Jones, A.~N. Gomez, L.~Kaiser, and I.~Polosukhin, ``Attention is all you need,'' 2023. [Online]. Available: \url{https://arxiv.org/abs/1706.03762}
\BIBentrySTDinterwordspacing

\bibitem{dw1000manual}
D.~Ltd., \emph{DW1000 User Manual Version 2.18}.

\end{thebibliography}




\end{document}